\documentstyle[12pt,epsfig,multirow]{article}

\newcommand{\be}{\begin{eqnarray}}
\newcommand{\ee}{\end{eqnarray}}

\newcommand{\ms}{\Delta m^2_{21}}
\newcommand{\ma}{\Delta m^2_{31}}

\newcommand{\sss}{\sin^2 \theta_{12}}
\newcommand{\sch}{\sin^2 \theta_{13}}

\newcommand{\sa}{\sin^2 \theta_{23}}

\newcommand{\mnu}{${\cal M}_\nu$}

\begin{document}

\thispagestyle{empty}
\begin{flushright}
\texttt{HRI-P-08-01-001}\\
\texttt{KEK-TH-1220}\\
\end{flushright}
\bigskip

\begin{center}
{\Large {\bf The $A_4$ flavor symmetry and neutrino phenomenology}}

\vspace{.5in}

{\bf Biswajoy Brahmachari$^{a,c}$, 
Sandhya Choubey$^b$, Manimala Mitra$^{b}$}
\vskip .5cm
$^a${\normalsize \it Department of Physics, 
Vidyasagar Evening College,}\\
{\normalsize \it 39, Sankar Ghosh Lane, Kolkata 700006, India}\\
\vskip .4cm
$^b${\normalsize \it Harish-Chandra Research Institute,} \\
{\normalsize \it Chhatnag Road, Jhunsi, Allahabad  211019, India}\\
\vskip .4cm
$^c${\normalsize \it Theory Division, KEK,} \\
{\normalsize \it Tsukuba, Ibaraki 305-0801, Japan}\\
\vskip 2cm

{\bf ABSTRACT}
\end{center}
\vskip 0.3cm

It has been shown that tribimaximal mixing can be 
obtained by some 
particular breaking pattern of the $A_4$ symmetry, 
wherein the extra $A_4$ triplet Higgs scalars pick up 
certain fixed vacuum expectation value (VEV) alignments. 
We have performed a detailed analysis of the 
different possible neutrino mass matrices within the 
framework of the $A_4$ model. 
We take into account all possible singlet and triplet 
Higgs scalars which leave the Lagrangian 
invariant under $A_4$. We break $A_4$ spontaneously, 
allowing the Higgs to take any VEV in general. 
We show that the neutrino mixing matrix deviates from tribimaximal, 
both due to the presence of the extra Higgs singlets, 
as well as from the 
deviation of the triplet Higgs VEV from its  
desired alignment, taken previously. 
We solve the 
eigenvalue problem for a variety of these illustrative  cases
and identify the ones where one obtains exact tribimaximal 
mixing. All such cases require fine-tuning. We show which neutrino mass matrices 
would be strongly disfavored 
by the current neutrino data. Finally, we study in detail the 
phenomenology of the remaining viable mass matrices and 
establish the deviation of the neutrino mixing from tribimaximal, 
both analytically as well as numerically.

\newpage

\section{Introduction}

Neutrinos have provided us with a window to physics beyond 
the Standard Model. A plethora of striking experimental 
results have propelled us to a juncture where we 
already know a great deal about the basic structure of the 
neutrino mixing matrix. Results from KamLAND \cite{kl} and 
solar neutrino experiments \cite{solar} 
can be best explained if $\sss = 0.32$, 
while atmospheric neutrino experiments \cite{atm} and 
results from K2K \cite{k2k} and MINOS \cite{minos} 
pick $\sa = 0.5$ as the best-fit solution. 
So far there has been no evidence for $\theta_{13}$ driven 
oscillations and hence $\theta_{13}$ is currently 
consistent with zero, with a $3\sigma$ upper bound of 
$\sin^2\theta_{13} < 0.05$ \cite{chooz,limits}. These results 
prod us to believe that the mixing matrix in the lepton sector 
could be of the tribimaximal (TBM) mixing form, first proposed 
by  Harrison, Perkins,
and Scott \cite{ref1,xingtbm},
\begin{equation}
U_{TBM}=\pmatrix{\sqrt{2 \over 3} & \sqrt{ 1 \over 3} & 0 \cr
-\sqrt{ 1 \over 6} & \sqrt{ 1 \over 3} & - \sqrt{ 1 \over 2} \cr
-\sqrt{ 1 \over 6} & \sqrt{ 1 \over 3} & \sqrt{ 1 \over 2}
}.
\label{eq:tbm}
\end{equation} 
It has been well known that for TBM mixing to 
exist, the neutrino mass matrix should be of the form
\be
{\cal M}_\nu = 
\pmatrix{A & B & B \cr
B & \frac{1}{2}(A+B+D) & \frac{1}{2}(A+B-D) \cr
B & \frac{1}{2}(A+B-D) & \frac{1}{2}(A+B+D) \cr
}~,
\label{eq:tbmmnu}
\ee
where $A=\frac{1}{3}(2\,m_1+m_2)$, 
$B=\frac{1}{3}(m_2-m_1)$ and $D=m_3$, where $m_1$, $m_2$ and $m_3$ 
are the neutrino masses. 
The current neutrino data already give very good measurement of 
mass squared differences, while the best limit on the 
absolute neutrino mass scale comes from cosmological data. 
We show in Table \ref{tab:current} the allowed 
neutrino oscillation parameters within their $3\sigma$ ranges.

\begin{table}[h]
\begin{center}
\begin{tabular}{c|c|c|c|c}
\hline\hline
$\ms$ & $\ma$ & $\sss$ & $\sa$ & $\sch$ \cr
\hline\hline
& & & &\cr
$(7.1-8.3)\times 10^{-5}$ eV$^2$ & $(2.0-2.8)\times 10^{-3}$ eV$^2$ 
& $0.26-0.40$ & $0.34-0.67$ & $<0.05$ \cr
& & & & \cr
        \hline
\end{tabular}
\caption{ \label{tab:current} 
The 
  3$\sigma$ allowed intervals (1 dof) for the
  three--flavor neutrino oscillation parameters from global data
  including solar, atmospheric, reactor (KamLAND and CHOOZ) and
  accelerator (K2K and MINOS) experiments.}
\end{center} 
\end{table}

Maximal $\theta_{23}$ and zero $\theta_{13}$ can be 
easily obtained if \mnu~ possesses $\mu-\tau$ 
exchange symmetry \cite{mutau} 
or the $L_\mu-L_\tau$ symmetry \cite{lmultau}. 
However, the solar mixing angle $\sss$ is not 
so easily predicted to be exactly 
1/3, as required for exact TBM mixing. 
While the current data points towards Eq. (\ref{eq:tbm}), 
better precision on the mixing angles are needed to 
really test the TBM ansatz and deviation from 
TBM \cite{deviation}.  
The mixing angles 
$\theta_{13}$ will be probed in the next generation 
long baseline and reactor experiments \cite{white}, while 
deviation of $\theta_{23}$ from maximality could be done 
in atmospheric neutrino experiments \cite{th23}. 
However, it is extremely crucial that one makes 
a very accurate measurement of $\theta_{12}$ in the future 
\cite{th12} in order to confirm TBM mixing. 

The challenge for model builders lies in explaining all features of the 
lepton mixing matrix together with the mass pattern  
of the neutrinos given in Table \ref{tab:current}. 
If indeed the neutrinos have TBM mixing, then one should 
be able to {\it naturally} generate the mass matrix given in 
Eq. (\ref{eq:tbmmnu}). 
Various symmetry 
groups explaining the flavor structure of the leptons have 
been invoked in the literature in order to accommodate the 
neutrino masses and mixing along with the charged leptons. 
In particular, 
the study of the non-Abelian discrete symmetry group $A_4$ 
has received considerable interest in the recent past 
\cite{ref2,maoriginal,ref3,ref5,babuhe,ma,others,tanimoto}. 
This group has been shown to 
successfully reproduce the TBM form of the 
neutrino mixing matrix, in the basis where the charged 
lepton mass matrix is diagonal \cite{ref5}. However, 
the authors of \cite{ref5} work in a very special 
framework where only one of the three possible singlet 
Higgs under $A_4$ is considered and for the $A_4$ triplet 
Higgs which contributes to the 
neutrino mass matrix, 
a particular vacuum alignment is taken\footnote{We give 
a brief overview of the $A_4$ model in section 2.}. 
In this framework,  
the mixing matrix emerges as apparently independent of the 
Yukawa couplings, the VEVs and the scale of $A_4$ 
breaking. Only the mass eigenstates depend on them. 

In this paper, we consider the most general scenario 
with all possible Higgs scalars that can be accommodated 
within this $A_4$ model. 
We give both approximate analytical solutions 
of the expected phenomenology as well as exact numerical 
results. We expound the conditions 
on VEVs and Yukawas 
needed for obtaining exact 
TBM mixing and show how the neutrino masses in those cases 
severely constrain them. We 
begin by showing that in the 
model considered in \cite{ref5}, one gets TBM mixing 
simply through the alignment of the $A_4$ triplet Higgs vev.
A concerted effort has been made in the literature to explain 
naturally this particular vacuum alignment needed for TBM 
mixing. The mass squared differences can be obtained if one has 
the vacuum expectation value (VEV) of an additional singlet Higgs. 
We show that to get the correct $\ms$ and $\ma$, 
the product of the VEV and Yukawa 
of this singlet is determined almost completely by the VEV 
and Yukawa of the triplet. This emerges as a further undesirable 
feature of the model, and one would need further explanation 
for this additional ``relative alignment'' 
between the product of VEVs and Yukawas 
of the triplet and the singlet. 

We allow for presence of an additional singlet Higgs, 
construct the neutrino
mass matrices and study their phenomenology. We take 
one, two and three singlet Higgs at a time and check which 
combinations produce viable neutrino mass matrices. In 
particular, we check which ones would produce TBM mixing 
and under what conditions. We obtain a few combinations
in addition to the one considered in \cite{ref5} which 
produce TBM mixing. We perturb these mass matrices by slightly 
shifting the VEVs and/or the Yukawas and study the deviation 
from TBM and the corresponding effect on the neutrino masses. 
We note that if 
just one singlet Higgs is allowed, the model of \cite{ref5} 
is the only viable model. We further show that in the 
simplest version of this model,  
one necessarily gets  
the normal mass hierarchy\footnote{We call this  
the neutrino mass {\it hierarchy}, though
what we mean is the {\it ordering} of the neutrino 
mass states.}. Inverted hierarchy can be possible 
if we have at least two or all three Higgs scalars with 
nonzero VEVs. 
We give predictions for the 
sum of the absolute neutrino masses ($m_t$), 
the effective mass that 
will be observed in neutrinoless double beta decay experiments
($\langle m_{ee} \rangle$)
and $m_\beta^2$ for tritium beta decay experiments. Finally, 
we perturb the VEV alignment for the triplet Higgs and 
study the deviation from TBM mixing. 

In section 2 we  give  a brief 
overview of the $A_4$ model considered. 
In section 3 we begin with detailed phenomenological 
analysis of the case where there is just one 
singlet Higgs under $A_4$.
We next increase the 
number of contributing singlet Higgs ,
give analytical 
and numerical results. 
In section 4 we study the impact of the misalignment 
of the VEVs of the triplet Higgs. We end in section 5
with our conclusions.

\section{Overview of the Model}

\begin{table}
\begin{center}
\begin{tabular}{|c|c|c|c|}
\hline
{\rm Lepton}& $SU(2)_L$ & $A_4$&\\
\hline
$l$&2&3&\\
$e^c$&1&1&\\
$\mu^c$&1&$1^{\prime\prime}$&\\
$\tau^c$&1&$1^\prime$&\\
\hline
Scalar&&&{\rm VEV}\\
\hline
$h_u$&2&1&$<h_u^0>$= $v_u$\\
$h_d$&2&1&$<h_d^0>$=$v_d$\\
$\phi_S$&1&3&$(v_S,v_S,v_S)$\\
$\phi_T$&1&3&$(v_T,0,0)$\\
$\xi$&1&1& $u$ \\
\hline
\hline
$\xi^\prime$&1&$1^\prime$& $u^\prime$\\
$\xi^{\prime \prime}$&1&$1^{\prime \prime}$& $u^{\prime \prime}$ \\
\hline
\end{tabular}
\caption{List of fermion and scalar fields used in this  model. Two
lower rows list additional singlets considered 
in the present work. In section 
4, we also allow for a different VEV alignment for $\phi_S$. 
}
\label{tab:particles}
\end{center}
\end{table}

Alternating group $A_n$  is a group of even permutations 
of $n$ objects. It  
is a subgroup of the permutation group $S_n$ and has  
$\frac{n!} {2} $ elements.
The non-Abelian group $A_4$ is the first 
alternating group which is not
a direct product of cyclic groups, and 
is isomorphic to the tetrahedral group $T_d$. 
The group $A_4$ has 12 elements, which can be written in terms of 
the generators of the 
group  $S$ and $T$. The generators of $A_4$ satisfy the relation 
\be
S^{2} = (ST)^{3} = (T)^{3}=1
\ee There are three 
one-dimensional irreducible representations of the group $A_4$ denoted as 
\be 
1 & S=1 & T=1\\
1' & S=1 & T=\omega^{2}\\
1'' &S=1 & T=\omega
\ee 
It is easy to check that there is no 
two-dimensional irreducible representation of this group.
The three-dimensional unitary representations of T and S are 
\be
T=\left(
\begin{array}{ccc}
1&0&0\\
0&\omega^2&0\\
0&0&\omega
\end{array}
\right),~~~~~~~~~~~~~~~~
S=\frac{1}{3}
\left(
\begin{array}{ccc}
-1&2&2\cr
2&-1&2\cr
2&2&-1
\end{array}
\right)~~~.
\label{ST}
\ee
where T has been chosen to be diagonal. 
We refer the readers to \cite{ref5} for 
a quick review of the $A_4$ group. Here we give multiplication 
rules for the singlet and triplet  representations,  for the sake of 
completeness. 
These multiplication rules correspond to the specific basis
of two generators $S,T$ of $A_4$.
 We have
\be
1\times 1 = 1,~~~~
1'\times 1'' = 1~~~~
3\times3 = 3 + 3_A + 1 + 1' + 1''~~
~.
\ee
For two triplets 
\be
a = (a_1,a_2,a_3),~~~~ b = (b_1,b_2,b_3)
\ee
one can write
\be
1\equiv (ab) &=& (a_1b_1 + a_2b_3 + a_3b_2)
\label{eq:one}\\
1'\equiv (ab)' &=& (a_3b_3 + a_1b_2 + a_2b_1)
\label{eq:onep}\\
1''\equiv (ab)'' &=& (a_2b_2 + a_1b_3 + a_3b_1)
\label{eq:onepp}
~.
\ee
Note that while 1 remains invariant under the exchange of 
the second and third elements of $a$ and $b$, $1'$ is 
symmetric under the exchange of the first and second elements while 
$1''$ is symmetric under the exchange of the first and third elements. 
\be
3\equiv (ab)_S &=& \frac{1}{3}\bigg(2a_1b_1 - a_2b_3 - a_3b_2,
2a_3b_3 - a_1b_2 - a_2b_1,2a_2b_2 - a_1b_3 - a_3b_1\bigg)
\label{eq:threes}\\
3_A\equiv (ab)_A &=& \frac{1}{3}\bigg(a_2b_3 - a_3b_2, 
a_1b_2 - a_2b_1, a_1b_3 - a_3b_1\bigg)
~.
\ee
We will be concerned with 
only $3$ and we can see that the first 
element here has 2-3 exchange symmetry,the  second element has 
1-2 exchange symmetry, while the third element has 
1-3 exchange symmetry. 
\footnote{There are no $3_A$ terms in the Lagrangian 
since it is antisymmetric and hence cannot be used for 
the neutrino mass matrix.}

The full particle content of the model we consider is shown 
in Table \ref{tab:particles}. 
There are five $SU(2)\otimes U_Y(1)$ Higgs 
singlets, three ($\xi$, $\xi^\prime$ and $\xi^{\prime\prime}$) 
of which 
are singlets under $A_4$ and two ($\phi_T$ and $\phi_S$)
of which transform as triplets. The standard 
model lepton doublets are assigned to the triplet representation 
of $A_4$, while the right handed charged leptons 
$e^c$, $\mu^c$ and $\tau^c$ are assumed to belong to 
the 1, $1^{\prime\prime}$ and $1^{\prime}$ representation 
respectively. The standard Higgs doublets $h_u$ and $h_d$ 
remain invariant under $A_4$.  The form of
the $A_4$ invariant Yukawa part of the Lagrangian is\footnote{
We assume that $\phi_S$ does not couple to charged leptons and
$\phi_T$ does not contribute to the Majorana mass matrix. 
These two additional features can be obtained from extra-dimensional
realization of the model, 
or from extra abelian symmetries \cite{ref5}.} 
\be
{\cal L}_{Y}=y_e e^c 
(\phi_T l)+y_\mu \mu^c (\phi_T l)'+
y_\tau \tau^c (\phi_T l)''+ x_a\xi (ll)+ x_a'\xi' (ll)''+ x_a''\xi'' (ll)'+
x_b (\phi_S ll)+h.c.+...
~
\label{eq:wl}
\ee
where, following \cite{ref5} we have used the compact notation, 
$y_e e^c (\phi_T l) \equiv y_e e^c (\phi_T l)h_d/\Lambda$,
$x_a \xi (ll) \equiv x_a \xi (lh_u lh_u)/\Lambda^2$ and so on, 
and $\Lambda$ is the cut-off scale of the theory. After 
the spontaneous breaking of $A_4$ followed
by $SU(2)_L \otimes U(1)_Y$, we get the mass terms 
for the charged leptons and neutrinos. 
Assuming the vacuum alignment
\be
\langle \phi_T \rangle=(v_T,0,0)
~,
\ee
the charged lepton mass matrix is given as 
\be
{\cal M}_l=\frac{v_d v_T}{\Lambda}\left(
\begin{array}{ccc}
y_e& 0& 0\\
0& y_\mu & 0 \\
0& 0& y_\tau 
\end{array}
\right)~~~,
\label{eq:mch}
\ee
Note that we could also obtain a diagonal charged lepton 
mass matrix even if we assume that $e^c$, $\mu^c$ and 
$\tau^c$ transform as $1''$, $1'$ and $1$, and 
$\langle \phi_T \rangle=(0,v_T,0)$ with appropriate change in the Yukawa Lagrangian. 
Similarly, $e^c$, $\mu^c$ and 
$\tau^c$ transform as $1'$, $1$ and $1''$, and 
$\langle \phi_T \rangle=(0,0,v_T)$ could give us the 
same ${\cal M}_l$. In what follows, we will assume that 
the ${\cal M}_l$ is of the form given in Eq. (\ref{eq:mch}).

In the most general case, 
where all three singlet Higgs 
as well as $\phi_S$ are present and we do not assume 
any particular vacuum alignment, the neutrino mass matrix looks like
\be
{\cal M}_\nu = m_0
\pmatrix{
a+2b_1/3 &c-b_3/3& d-b_2/3 \cr
c-b_3/3  & d+2b_2/3 & a-b_1/3 \cr
d-b_2/3& a-b_1/3& c+2b_3/3 \cr
}
~,
\label{eq:generalmnu}
\ee
where 
$m_0 = \frac {v_u^2}{\Lambda}$
$b_i=2 x_b \frac{v_{S_i}}{\Lambda}$, $a=2 x_a\frac{u}{\Lambda}$,
$c=2 x_a''\frac{u''}{\Lambda}$ and  $d=2 x_a' \frac {u'}{\Lambda}$
and we have written the VEVs as 
\be
\langle \phi_S \rangle = (v_{S_1},v_{S_2},v_{S_3}),~~~
\langle \xi \rangle=u,~~~
\langle \xi' \rangle=u',~~~
\langle \xi'' \rangle=u'',~~~
\langle h_{u,d}\rangle= v_{u,d}
~.
\label{eq:vev}
\ee

\section{Number of Singlet Higgs and their VEVs}

In this section we work under the assumption that 
the triplet Higgs $\phi_S$ has VEVs along the direction
\be
\langle \phi_S \rangle &=& (v_S,v_S,v_S)
~.
\ee
This produces the neutrino mass matrix  
\be
{\cal M}_\nu = 
m_0
\pmatrix{
a+2b/3 &c-b/3& d-b/3 \cr
c-b/3  & d+2b/3 & a-b/3 \cr
d-b/3& a-b/3& c+2b/3 \cr
}
~,
\label{eq:phisalignedmnu}
\ee
where $b=2 x_b \frac{v_{S}}{\Lambda}$. In the following 
we discuss the phenomenology of the different forms of 
${\cal M}_\nu$ possible as we change the number of singlet 
Higgs or put their VEVs to zero. We assume that \mnu~
is real. 

\subsection{No $A_4$ Singlet Higgs}

If there were no singlet Higgs, or if the VEV of all three of 
them were zero, one would get the neutrino mass matrix
\be
{\cal M}_\nu = 
m_0
\pmatrix{
2b/3 &-b/3& -b/3 \cr
-b/3  & 2b/3 & -b/3 \cr
-b/3& -b/3& 2b/3 \cr
}
~.
\label{eq:phisalignedmnu0}
\ee
On diagonalizing this one obtains the eigenvalues 
\be
m_1 = m_0\,b,~~~m_2=0,~~~m_3= m_0\,b
\ee
and the mixing matrix
\be
U = 
\pmatrix{\sqrt{2 \over 3} & \sqrt{ 1 \over 3} & 0 \cr
-\sqrt{ 1 \over 6} & \sqrt{ 1 \over 3} & - \sqrt{ 1 \over 2} \cr
-\sqrt{ 1 \over 6} & \sqrt{ 1 \over 3} & \sqrt{ 1 \over 2}
}.
\ee
Therefore, we can see that the tribimaximal 
pattern of the mixing matrix is coming directly from the 
term containing the triplet Higgs $\phi_S$ and does not depend 
on the terms containing the singlet Higgs scalars\footnote{Note that 
the mixing pattern does not depend explicitly even on the 
VEV of $\phi_S$.}. 
However, in the absence of the singlet Higgs contributions to 
\mnu, the ordering of the neutrino masses turn out to be very wrong. 
In this case, $\ms = -b^2 \,m_0^2$ 
and $\ma=0$, in stark disagreement with 
the oscillation data.

\subsection{Only one $A_4$ Singlet Higgs}

 \begin{table}
        \begin{center}
        
        \begin{tabular}{|l|c|c|c|}
        \hline
            Higgs & Neutrino mass matrix & Eigenvalues & Mixing Matrix \\
            \hline
           
            $\xi$ &  
$
m_0\left(
\begin{array}{ccc}
a+\frac{2b}{3} &-\frac{b}{3}& -\frac{b}{3}\\
-\frac{b}{3}  & \frac{2b}{3} & a-\frac{b}{3} \\
-\frac{b}{3}& a-\frac{b}{3}& \frac{2b}{3} 
\end{array}
\right)
$
&

$

\left(\begin{array}{c}
m_0(a+b),\\
m_0a,\\
m_0(b-a)
\end{array}
\right)
$

&   
 
$
\left(\begin{array}{ccc}
\sqrt{\frac{2}{3}} & \frac{1}{\sqrt{3}} & 0 \\
-\frac{1}{\sqrt{6}} & \frac{1}{\sqrt{3}} & -\frac{1}{\sqrt{2}} \\
-\frac{1}{\sqrt{6}} & \frac{1}{\sqrt{3}} & \frac{1}{\sqrt{2}} 
\end{array}
\right)
$

\\&&&\\
\hline

$\xi''$
&
$ m_0\left(
\begin{array}{ccc}
\frac{2b}{3} &c-\frac{b}{3}& -\frac{b}{3}\\
c-\frac{b}{3}  & \frac{2b}{3} & -\frac{b}{3} \\
-\frac{b}{3}& -\frac{b}{3}& c+\frac{2b}{3} 
\end{array}
\right)
$
&
$
\left(\begin{array}{c}
m_0(c+b),\\
m_0c,\\
m_0(b-c) 
\end{array}
\right)
$
&

$
\pmatrix{
-\frac{1}{\sqrt{6}} & \frac{1}{\sqrt{3}} & -\frac{1}{\sqrt{2}} \cr
-\frac{1}{\sqrt{6}} & \frac{1}{\sqrt{3}} &  \frac{1}{\sqrt{2}} \cr
\sqrt{\frac{2}{3}} & \frac{1}{\sqrt{3}} & 0 \cr
}
$

\\&&&\\
\hline
$\xi'$
&
$ m_0\left(
\begin{array}{ccc}
\frac{2b}{3} &-\frac{b}{3}& d-\frac{b}{3}\\
-\frac{b}{3}  & d+\frac{2b}{3} & -\frac{b}{3} \\
d-\frac{b}{3}& -\frac{b}{3}& \frac{2b}{3} 
\end{array}
\right)
$
&
$
\left(\begin{array}{c}
m_0(d+b),\\
m_0d,\\
m_0(b-d) 
\end{array}
\right)
$
&

$
\pmatrix{
-\frac{1}{\sqrt{6}} & \frac{1}{\sqrt{3}} & -\frac{1}{\sqrt{2}} \cr
\sqrt{\frac{2}{3}} & \frac{1}{\sqrt{3}} & 0 \cr
-\frac{1}{\sqrt{6}} & \frac{1}{\sqrt{3}} &  \frac{1}{\sqrt{2}} \cr
}
$

\\&&&\\ \hline
\end{tabular}
\end{center}
\caption{The mass matrix taking 
one singlet at a time, its mass eigenvalues,
and its mixing matrix.}
\label{tab:onehiggs}
\end{table}

\begin{figure}
\centering
\includegraphics[width = 18cm,height=16cm,angle = 0]{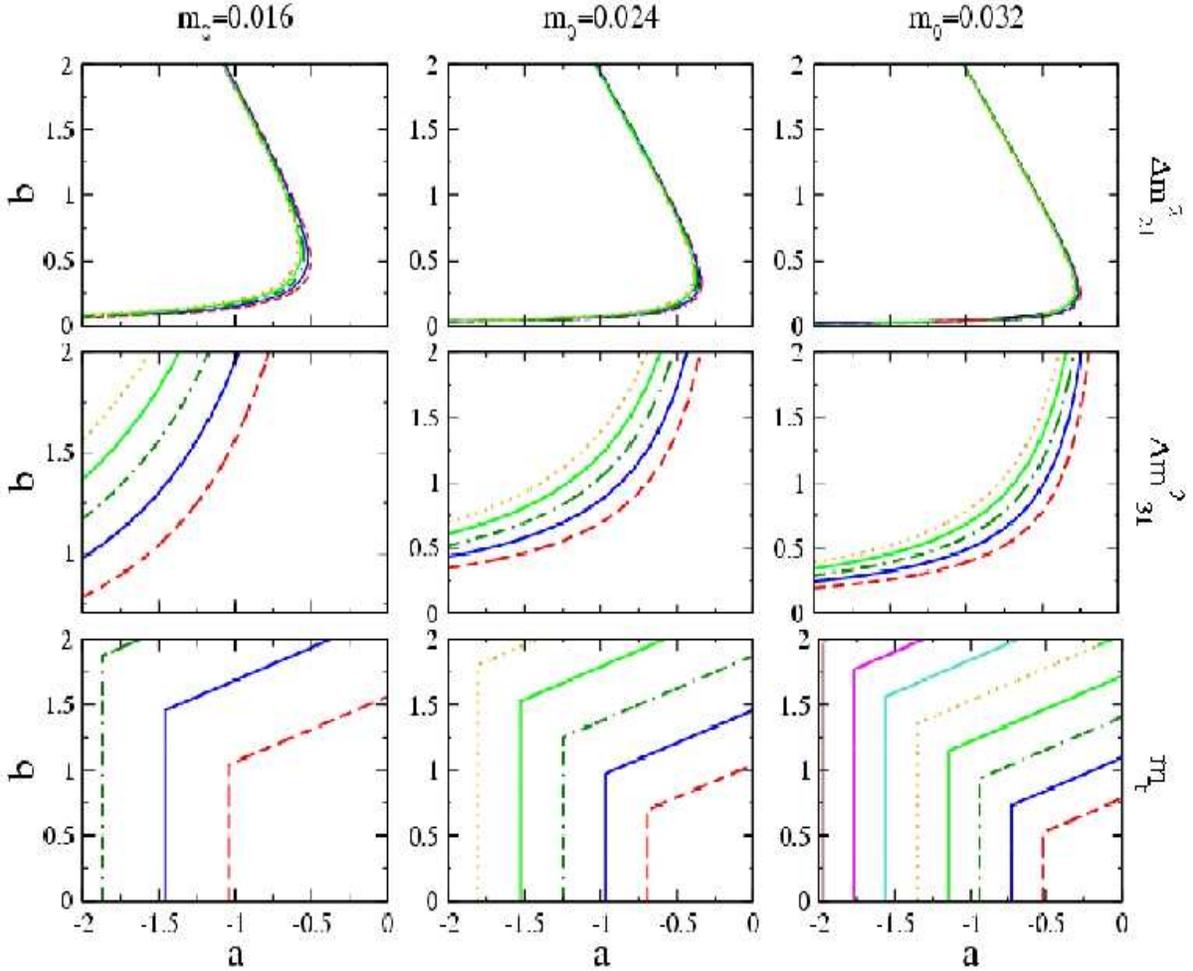}
\caption{Contour plot of 
$\Delta m^{2}_{21}, \Delta m^{2}_{31} $ and sum of 
absolute neutrino masses $m_{t}$  in the $a-b$ plane, 
for three different values of $m_{0}$ 
($m_{0}$=0.016, $m_{0}$=0.024 
and $m_{0}$=0.032) for the case with $\xi$ and $\phi_S$. 
The first row shows contour plots for different values 
of $\Delta m^{2}_{21}$ (in $10^{-5}$ eV$^{2}$), with 
red dashed lines for 6.5
blue solid lines for 7.1, green dashed 
lines for 7.7, green solid 
lines for 8.3, and orange dotted 
lines for 8.9. The second row shows contour 
plots for different values 
of $\Delta m^{2}_{31}$ (in $10^{-3}$ eV$^{2}$), with red 
dashed lines for 1.6,
blue solid 
lines for  2.0, green 
dashed lines for 2.4, 
green solid lines for 2.8, and  
orange dotted lines for 3.2. The third 
row shows contour plots for $m_t$ 
(in eV), with red dashed lines for 0.05, 
blue solid lines for 0.07, 
 green dashed lines for 0.09,
green solid lines for 1.1,
orange dotted lines for 1.3,
turquoise solid lines for 1.4,
magenta solid lines for 1.6,
and brown solid lines for 1.8.
}
\label{fig:cont}
\end{figure}

\begin{figure}
\centering
\includegraphics[width=10.0cm,height=7.0cm,angle=0]{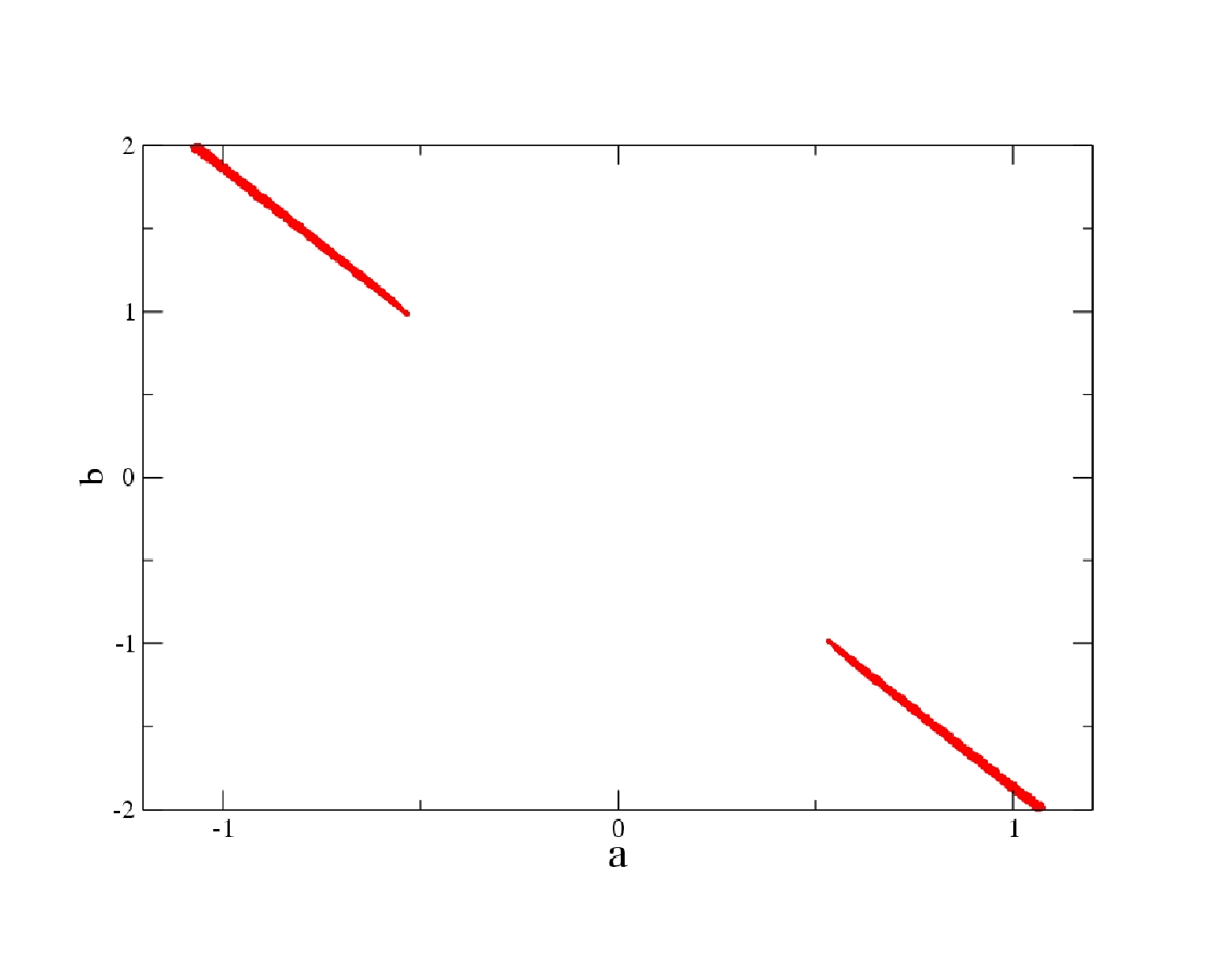}
\caption{Scatter plot showing regions in $a-b$ parameter 
space for the model considered in \cite{ref5}, 
which are compatible with the 
current $3\sigma$ allowed range of values of the mass squared 
differences. The parameter $m_0$ is allowed to vary freely. 
}
\label{fig:ab}
\end{figure}

If we take only one $A_4$ singlet Higgs at a time then there 
are three possibilities. The resulting mass matrices are shown 
in column 2 of Table \ref{tab:onehiggs}. One could get exactly 
the same situation with three singlet Higgs and demanding that the 
VEV of two of them are zero while that of the third is nonzero. 
The ${\cal M}_\nu$ given in Table \ref{tab:onehiggs} can be 
exactly diagonalized and the eigenvalues and eigenvectors 
are shown in column 3 and 4 of the Table, respectively. 
One can see that only the case where $\xi$ is present gives 
rise to a viable mixing matrix, which is  exactly 
tribimaximal \cite{ref5}.  
One can check that only the first case has the 
form for \mnu~ given in Eq. (\ref{eq:tbmmnu}). 
 Note that each of the \mnu~ given in 
Table \ref{tab:onehiggs} possesses an
$S_2$ symmetry, which reflects the symmetry of  
Eqs. (\ref{eq:one}-\ref{eq:onepp})\footnote{Since we 
have assumed the vacuum alignment  
$\langle \phi_S \rangle = (v_S,v_S,v_S)$, 
the $b$ terms of \mnu~ are such that it is symmetric 
under all the three exchange symmetries.}.  
While the case with $\xi$ exhibits the $\mu-\tau$ exchange 
symmetry, the one with $\xi''$ remains invariant 
under $e-\mu$ permutation and the one with $\xi{^\prime}$
under $e-\tau$ permutation. This would necessarily demand that 
while for the first case $\theta_{23}$ would be maximal and 
$\theta_{13}=0$, for the second and third cases $\theta_{23}$ 
would be either $90^\circ$ or 0 respectively, and 
$\theta_{13}$ maximal. 

Since only the case with $\xi$ reproduces the correct form for the 
mixing matrix, we do not discuss the remaining two 
cases any further. 
This was the model presented by Altarelli and Feruglio 
in their seminal paper \cite{ref5}. The mass squared differences
in this case are
\be
\ms = (-b^2 - 2ab)m_0^2,~~~\ma=-4\,a\,b\,m_0^2
~, 
\ee
where $m_0=v_u^2/\Lambda$.  
Since it is now known at more than $6\sigma$ C.L. that $\ms >0$ 
\cite{limits}, we have the condition that $-2ab>b^2$. Since 
$b^2$ is a positive definite quantity the above relation 
implies that $-2ab >0$ , which can happen if and only if 
$sgn(a) \neq sgn(b)$. Inserting this condition into the 
expression for $\ma$ gives us $\ma > 0$ necessarily in this 
model. Therefore, inverted neutrino mass 
hierarchy is impossible to get in the framework of the
simplest version of the model proposed by 
Altarelli and Feruglio \cite{ref5}.

The sum of the absolute neutrino masses, 
effective mass in neutrinoless double beta decay and
prediction for tritium beta decay are given 
respectively as 
\be
m_t = |m_1|+|m_2|+|m_3|,~~~~\langle m_{ee} \rangle  = m_0(a+ 2b/3),~~~~
m_\beta^2 = m_0^2\bigg(a^2  +  \frac{4ab}{3} +  \frac{2b^2}{3}\bigg) 
~.
\ee
In Fig. \ref{fig:cont} we show the contours for the 
observables $\ms$, $\ma$ and $m_t$ in the $a-b$ plane,
for three different fixed values of $m_0$. The details 
of the figure and description of the different lines can be 
found in the caption of the figure.

In Fig. \ref{fig:ab} we present a scatter 
plot showing the points in the $a-b$ parameter space which 
are compatible with the 
current $3\sigma$ allowed range of the 
mass squared differences. 
We have allowed $m_0$ to vary freely and taken a projection of 
all allowed points in the $a-b$ plane. 
Note that 
while $a$ is related to the VEV of the singlet $\xi$, 
$b$ is given in terms of the VEVs of the triplet 
$\phi_S$. We reiterate the point mentioned before 
that the TBM form for the mixing matrix 
comes solely from the vacuum alignment of $\phi_S$ and 
$\xi$ is not needed for that. The singlet 
$\xi$ is necessary only for producing the correct 
values of $\ms$ and $\ma$. However, we can note 
from Fig. \ref{fig:ab} that for a given value of 
$a$ needed to obtain the 
right mass squared differences, the value of $b$ 
is almost fixed. In fact, we can calculate the relation 
between $a$ and $b$ by looking at the ratio
$
\frac{\ms}{\ma} = \frac{-b^2-2ab}{-4ab} \simeq 0.03
$,
where 0.03 on the right-hand-side (RHS) is the current experimental value. 
This gives us the relation $b\simeq -1.88 a$. 
Therefore, not only does one need 
the alignment $ \langle \phi_S \rangle= (v_S,v_S,v_S)$ 
to get TBM, one also needs a particular relation between  
the product of Yukawa couplings and 
VEVs of $\phi_S$ and $\xi$ in order to reproduce the 
correct phenomenology. This appears to be rather contrived. 
Even if one 
includes the $3\sigma$ uncertainties on $\ms$ and $\ma$, 
$|b|$ is fine tuned to $|a|$ within a factor of about $10^{-2}$. 

\subsection{Two $A_4$ Singlet Higgs}

\begin{table}
        \begin{center}
        \begin{tabular}{|l|c|c|c|}
        \hline
            Higgs & Neutrino mass matrix & Eigenvalues & Eigenvectors \\
            \hline
$\xi$,$\xi^{\prime\prime}$
&
$ m_0\left(
\begin{array}{ccc}
a+\frac{2b}{3} &c-\frac{b}{3}& -\frac{b}{3}\\
c-\frac{b}{3}  & \frac{2b}{3} & a-\frac{b}{3} \\
-\frac{b}{3}& a-\frac{b}{3}& c+\frac{2b}{3} 
\end{array}
\right)
$
&
a=c;
&
$ \left(
\begin{array}{ccc}
-\frac{1}{\sqrt{6}} & \frac{1}{\sqrt{3}} & -\frac{1}{\sqrt{2}} \\
\sqrt{\frac{2}{3}} & \frac{1}{\sqrt{3}} & 0 \\
-\frac{1}{\sqrt{6}} & \frac{1}{\sqrt{3}} & \frac{1}{\sqrt{2}} \\
\end{array}
\right)
$
\\
&&
 $
\left(\begin{array}{c}

m_0(b-a),\\
2m_0a,\\
m_0(a+b) 
\end{array}
\right)
$
&
\\ \hline
$\xi$,$\xi^{\prime}$ 
&
$m_0
 \left(
\begin{array}{ccc}
a+\frac{2b}{3} &-\frac{b}{3}& d-\frac{b}{3}\\
-\frac{b}{3}  & d+\frac{2b}{3} & a-\frac{b}{3} \\
d-\frac{b}{3}& a-\frac{b}{3}& \frac{2b}{3} 
\end{array}
\right)
$
&a=d;
&
$ \left(
\begin{array}{ccc}
-\frac{1}{\sqrt{6}} & \frac{1}{\sqrt{3}} & -\frac{1}{\sqrt{2}} \\
-\frac{1}{\sqrt{6}} & \frac{1}{\sqrt{3}} & \frac{1}{\sqrt{2}} \\
\sqrt{\frac{2}{3}} & \frac{1}{\sqrt{3}} & 0 \\
\end{array}
\right)
$ 
\\
&&
$
\left(\begin{array}{c}
m_0(b-a),\\
2m_0a,\\
m_0(a+b) 
\end{array}
\right)
$
&
\\ \hline
$\xi^\prime$,$\xi^{\prime \prime}$
&
$m_0
 \left(
\begin{array}{ccc}
\frac{2b}{3} &c-\frac{b}{3}& d-\frac{b}{3}\\
c-\frac{b}{3}  & d+\frac{2b}{3} & -\frac{b}{3} \\
d-\frac{b}{3}& -\frac{b}{3}& c+\frac{2b}{3} 
\end{array}
\right)
$
&c=d;
&
$\left(
\begin{array}{ccc}
\sqrt{\frac{2}{3}} & \frac{1}{\sqrt{3}} & 0 \\
-\frac{1}{\sqrt{6}} & \frac{1}{\sqrt{3}} & -\frac{1}{\sqrt{2}} \\
-\frac{1}{\sqrt{6}}& \frac{1}{\sqrt{3}} & \frac{1}{\sqrt{2}} \\
\end{array}
\right)
$
\\
&&
$
\left(
\begin{array}{c}
m_0(b-c),\\
2m_0c,\\
m_0(c+b)
\end{array}
\right)
$
&
\\ \cline{3-4}
&&
$c=d+\epsilon$;
&
$\left(
\begin{array}{ccc}\\
\sqrt{\frac{2}{3}} & \frac{1}{\sqrt{3}} & -\frac{\epsilon}{2\sqrt{2}d}\\
-\frac{1}{\sqrt{6}} -\frac{\sqrt{3}\epsilon}{4\sqrt{2}d} &
 \frac{1}{\sqrt{3}} &  -\frac{1}{\sqrt{2}} + 
\frac{\epsilon}{4\sqrt{2}d}\\
-\frac{1}{\sqrt{6}} + \frac{\sqrt{3}\epsilon}{4\sqrt{2}d} &
 \frac{1}{\sqrt{3}} &  \frac{1}{\sqrt{2}} + 
\frac{\epsilon}{4\sqrt{2}d}\\
\end{array}
\right)
$    
\\&&
$
\left(
\begin{array}{c}
m_0(b-d-\frac{\epsilon}{2}),\\
m_0(2d+\epsilon),\\
m_0(d+b+\frac{\epsilon}{2})
\end{array}
\right)
$
&
\\\hline
\end{tabular}
\end{center}
\caption{Here we take two singlets at a time, and analytically display 
eigenvalues and eigenvectors of the neutrino mass matrix.}
\label{tab:twohiggs}
\end{table}

\begin{figure}
\centering
\includegraphics[width = 19cm,height=18cm,angle = 0]{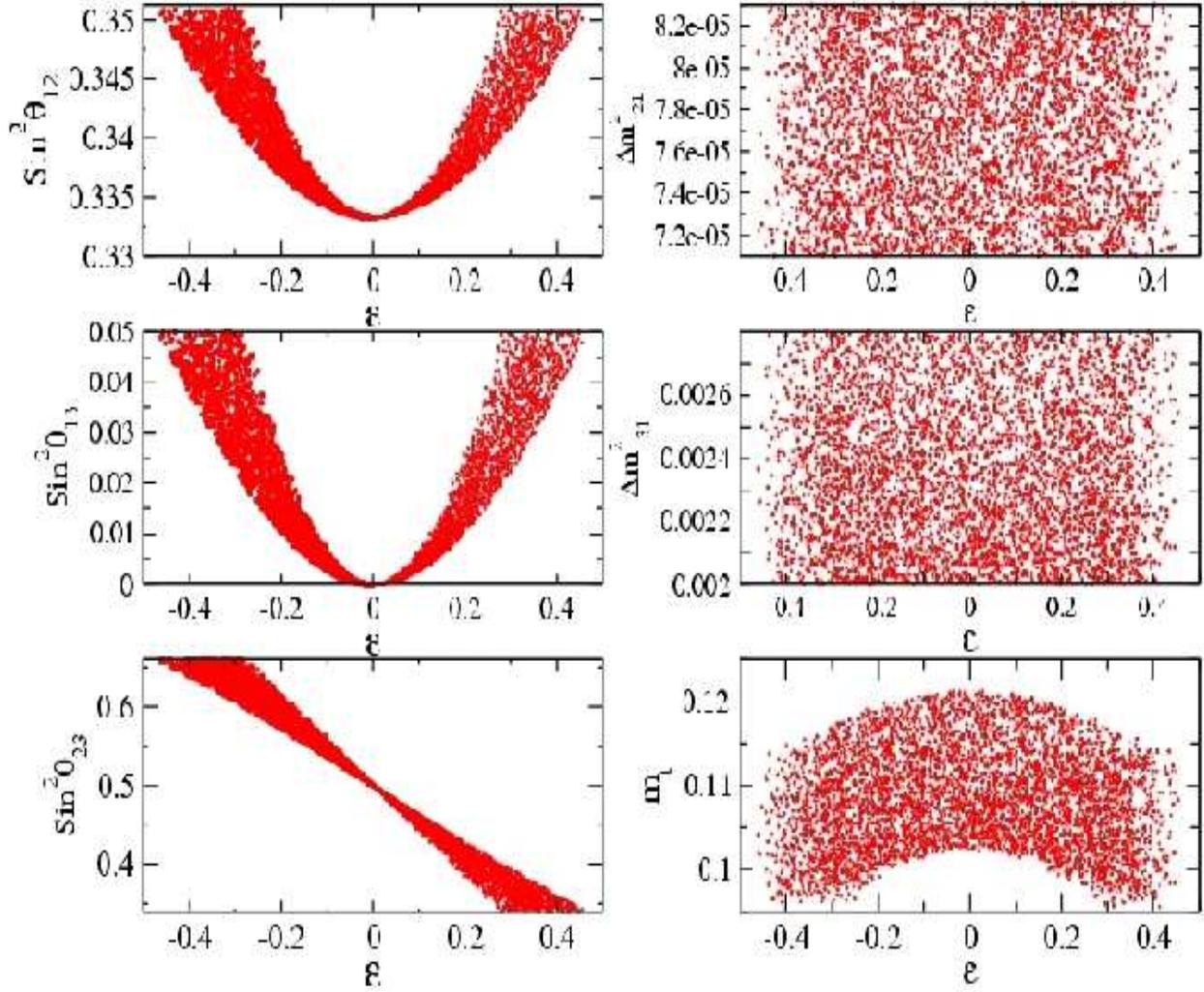} 
\caption{The left panels show 
$\sin^{2}\theta_{12}$, $\sin^{2}\theta_{13}$ 
and $\sin^{2}\theta_{23}$ vs $\epsilon$   
and the right panels show $\Delta m^2_{21}$, $\Delta m^{2}_{31}$ 
and $m_{t}$ vs $\epsilon$ respectively. 
Here $\xi^\prime$ and $\xi^{\prime \prime}$ acquire VEVs. 
The other parameters $c$, $b$ and $m_0$ are allowed 
to vary freely. 
}
\label{fig:twohiggs}
\end{figure}

\begin{figure}
\centering
\includegraphics[width = 19cm,height=18cm,angle = 0]{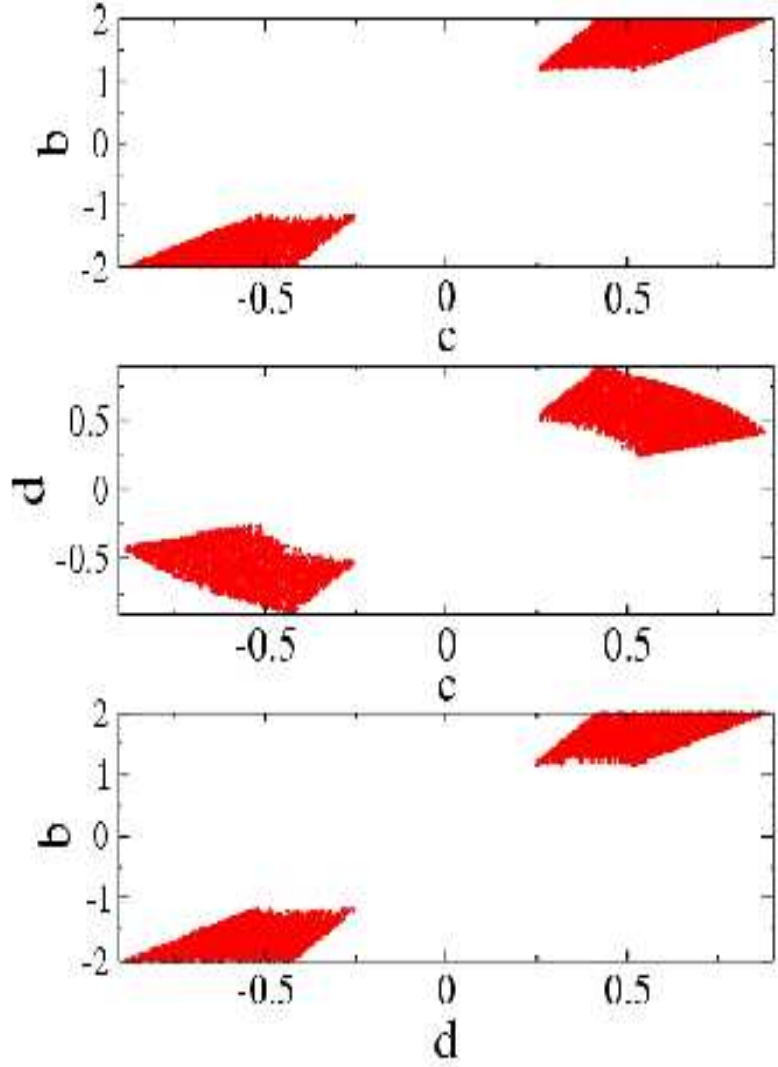}
\caption{Scatter plot showing the $3\sigma$ allowed regions for the
$b-c-d$ parameters for the case where 
$\xi^\prime$ and $\xi^{\prime \prime}$ 
acquire VEVs.
The top, middle and lower 
panels show the allowed points projected on 
the $c-b$, $c-d$ and $d-b$ plane, respectively. 
The parameter $m_0$ was allowed to take any value. Here 
we have assumed normal hierarchy.
}
\label{fig:twohiggs_param}
\end{figure}

\begin{figure}
\centering
\includegraphics[width = 19cm,height=18cm,angle = 0]{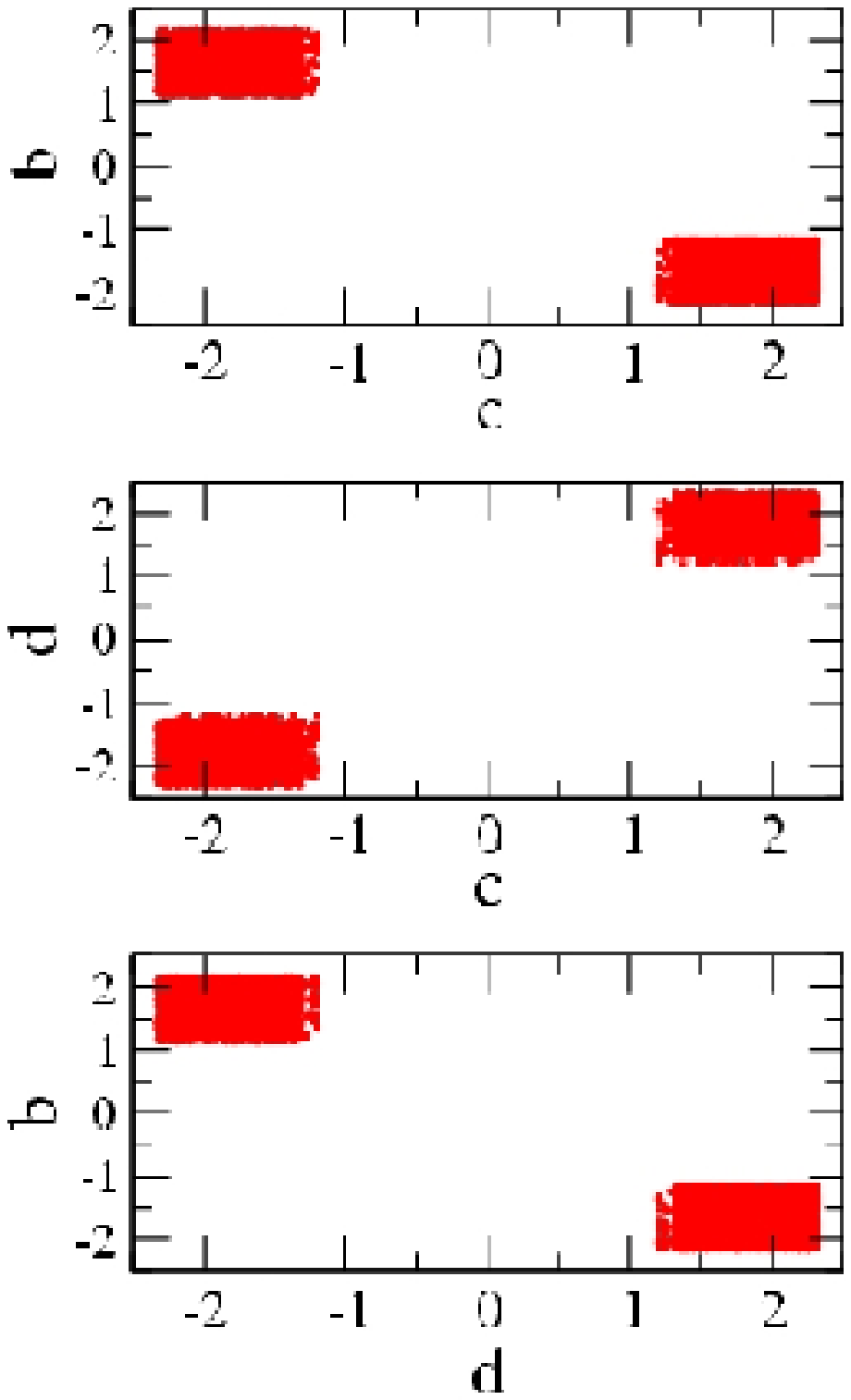}
\caption{\label{fig:twohiggsih}
Same as in Fig. \ref{fig:twohiggs_param} but for inverted 
hierarchy.
}
\end{figure}

If we take two singlet Higgs belonging to two different singlet 
representations and allow for nonzero 
VEVs for them, then the \mnu~ obtained for the 
three possible cases are shown in 
the first three rows of Table \ref{tab:twohiggs}. 
One can again see that of the three possible combinations, 
only the $\xi^\prime$, $\xi^{\prime\prime}$ combination 
gives a viable TBM matrix. The other two mass matrices exhibit 
$e-\tau$ ($\xi$, $\xi^{\prime\prime}$)and $e-\mu$ 
($\xi$, $\xi^{\prime}$) symmetry respectively and are 
ruled out. Note that we have chosen $a=c$ for the 
$\xi$, $\xi^{\prime\prime}$ combination, $a=d$ for 
the $\xi$, $\xi^{\prime}$ combination and 
$c=d$ for the $\xi^\prime$, $\xi^{\prime\prime}$ combination 
for the results given in Table \ref{tab:twohiggs}. This is a 
reasonable assumption to make since the phenomenology of the 
three cases does not change drastically unless the VEVs of 
the singlet Higgs vary by a huge amount. 
In particular, 
by changing the relative magnitude of the VEVs, we 
do not expect the structure of the mixing matrix for the first 
two rows of Table \ref{tab:twohiggs} 
to change so much so that they could be allowed by the 
current data. In the limit that $c=d$, it is not hard 
to appreciate that the resultant matrix with $\xi'$ and 
$\xi^{\prime\prime}$ would exhibit $\mu-\tau$ symmetry, though 
the $\xi^{\prime}$ and $\xi^{\prime\prime}$ terms alone have  
$e-\tau$ and $e-\mu$ symmetry respectively.

Since the $\xi^\prime$, $\xi^{\prime\prime}$ combination 
is the only one which gives exact TBM mixing 
in the approximation that $c=d$, 
we perform a detailed analysis only for this case. 
Putting $c=d$ is again contrived and would 
also lead to a certain fixed relation between them 
and $b$, as in the only $\xi$ case. This would mean 
additional fine tuning of the paramaters, unless explained 
by symmetry arguments. Hence, 
we allow the two VEVs to differ from 
each other so that $c=d+\epsilon$. If $\epsilon$ is small 
we can solve the eigenvalue problem keeping only the first 
order terms in $\epsilon$. The results for this case are 
shown in the final row of the Table \ref{tab:twohiggs}. 
The deviation of the mixing angles from their TBM values can be
seen to be
\be
D_{12} \simeq 0,~~~~D_{23} \simeq -\frac{\epsilon}{4d},~~~~
U_{e3} \simeq -\frac{\epsilon}{2\sqrt{2}d}
~, 
\label{eq:twohiggs}
\ee
where $D_{12} = \sss - 1/3$ and $D_{23} = \sa -1/2$.  
We show in the 
left hand panels of Fig. \ref{fig:twohiggs} 
the mixing angles $\sss$ (upper panel), 
$\sch$ (middle panel) and $\sa$ (lower panel) 
as a function of $\epsilon$. We vary $\epsilon$ 
from large negative to large positive values and 
solve the exact eigenvalue problem numerically 
allowing the other parameters, $m_0$, $b$ and $d$, to vary 
freely. 
For $\epsilon=0$ of course we get TBM mixing as expected. For 
very small values of $\epsilon$, the deviation of the mixing 
angles from their TBM values is reproduced well by the 
approximate expressions given in Eq. (\ref{eq:twohiggs}).
For large $\epsilon$ of course the approximate expressions fail and 
we have significant deviation from TBM. The values 
of $\sss$ and $\sch$ increase 
very fast with $\epsilon$, while $\sa$ decreases with it. 
We have only showed the scatter plots up to the current 
$3\sigma$ allowed 
ranges for the mixing angles. 
The mass dependent observables can be calculated upto 
first order in $\epsilon$ as
\be
\ms \simeq m_0^2\,(b+d+\frac{\epsilon}{2})(3d-b+\frac{3\epsilon}{2}),~~~~
\ma \simeq (4bd + 2b\epsilon)m_0^2
~,
\ee
\be
\langle m_{ee} \rangle  =  m_0\frac{2b}{3},~~~~
m_t = \sum_i |m_i|,~~~~
m_\beta^2 \simeq
m_0^2(\frac{2b^2}{3}+2d^2 - \frac{4bd}{3}+2d\epsilon -\frac{2b\epsilon}{3})
~.
\ee
Of course, epsilon can be larger and we show numerical results for 
those cases. 

\subsubsection{Normal Hierarchy}

The right panels of Fig. \ref{fig:twohiggs} show 
$\ms$ (upper panel), 
$\ma$ (middle panel) and $m_t$ (lower panel) 
as a function of $\epsilon$ assuming 
$m_1 < m_2 \ll m_3$. 
We show $\ms$ and 
$\ma$ only within their $3\sigma$ allowed 
range. We notice that while $\ms$ and $\ma$ are hardly 
constrained by $\epsilon$, there appears to some mild 
dependence of $m_t$ on it. 

Fig. \ref{fig:twohiggs_param} gives the scatter plots 
showing the allowed parameter regions for this case. 
The upper, middle and lower 
panels show the allowed points projected on the 
$b-c$, $d-c$ and $b-d$ plane, respectively. 

\subsubsection{Inverted Hierarchy}

For this case it is possible to obtain even inverted
hierarchy. We show in Fig. \ref{fig:twohiggsih} 
the scatter plots showing the allowed parameter 
regions for inverted hierarchy. We have allowed 
$m_0$ to vary freely and show the 
allowed points projected on the $c-b$, $c-d$ and $d-b$ planes. 
One can check that only for the points 
appearing in this plot, $m_3 < m_1 < m_2$. We reiterate 
that these points also satisfy the $3\sigma$ allowed 
oscillation parameter ranges given in Table \ref{tab:current}.

\subsection{Three $A_4$ Singlet Higgs}

  \begin{table}
        \begin{center}
        \begin{tabular}{|l|c|c|c|}
        \hline
            Higgs & Neutrino mass matrix & Eigenvalues & Eigenvectors \\
            \hline
$\begin{array}{c}
\xi,\\
\xi^\prime,\\
\xi^{\prime \prime}
\end{array}$
&
$
m_0
 \left(
\begin{array}{ccc}
a+\frac{2b}{3} &c-\frac{b}{3}& d-\frac{b}{3}\\
c-\frac{b}{3}  & d+\frac{2b}{3} & a-\frac{b}{3} \\
d-\frac{b}{3}& a-\frac{b}{3}& c+\frac{2b}{3} 
\end{array}
\right)
$
& $a=c=d$;
&
   $ \left(
\begin{array}{ccc}
-\frac{1}{\sqrt{6}} & \frac{1}{\sqrt{3}} & -\frac{1}{\sqrt{2}} \\
-\frac{1}{\sqrt{6}} & \frac{1}{\sqrt{3}} & \frac{1}{\sqrt{2}} \\
\sqrt{\frac{2}{3}} & \frac{1}{\sqrt{3}} & 0 \\
\end{array}
\right)
$ 
\\&&    
$
\left(\begin{array}{c}
m_0b,\\
3m_0a,\\
m_0b
\end{array}
\right)
$
&
\\ \cline{3-4}
&&
$a \neq c= d$;
&
$\left(
\begin{array}{ccc}
\sqrt{\frac{2}{3}} & \frac{1}{\sqrt{3}} & 0 \\
-\frac{1}{\sqrt{6}} & \frac{1}{\sqrt{3}} & -\frac{1}{\sqrt{2}} \\
-\frac{1}{\sqrt{6}}& \frac{1}{\sqrt{3}} & \frac{1}{\sqrt{2}} \\
\end{array}
\right)
$
\\&&
$
\left(\begin{array}{c}
m_0(a+b-c),\\
m_0(a+2c),\\
m_0(b+c-a)
\end{array}
\right)
$
&
\\ \cline{3-4}
&&
$a=c \neq d$;
&

$ \left(
\begin{array}{ccc}
-\frac{1}{\sqrt{6}} & \frac{1}{\sqrt{3}} & -\frac{1}{\sqrt{2}} \\
\sqrt{\frac{2}{3}} & \frac{1}{\sqrt{3}} & 0 \\
-\frac{1}{\sqrt{6}} & \frac{1}{\sqrt{3}} & \frac{1}{\sqrt{2}} \\
\end{array}
\right)
$

\\&&
$
\left(\begin{array}{c}
m_0(b+d-a),\\
m_0(2a+d),\\
m_0(a+b-d) 
\end{array}
\right)
$
&
\\\cline{3-4}
&&
$a=d \neq c$;
&
$ \left(
\begin{array}{ccc}
-\frac{1}{\sqrt{6}} & \frac{1}{\sqrt{3}} & -\frac{1}{\sqrt{2}} \\
-\frac{1}{\sqrt{6}} & \frac{1}{\sqrt{3}} & \frac{1}{\sqrt{2}} \\
\sqrt{\frac{2}{3}} & \frac{1}{\sqrt{3}} & 0 \\
\end{array}
\right)
$ 

\\&&
$
\left(\begin{array}{c}
m_0(b+c-a)\\
m_0(2a+c),\\
m_0(a+b-c) 
\end{array}
\right)
$
&
\\\cline{3-4}
&&
$a\neq c=d+\epsilon$;
&
$ \left(
\begin{array}{ccc}
\sqrt{\frac{2}{3}} & \frac{1}{\sqrt{3}} & a_4\\
-\frac{1}{\sqrt{6}}+a_2 & \frac{1}{\sqrt{3}} & -\frac{1}{\sqrt{2}}-a_5\\
-\frac{1}{\sqrt{6}}+a_3& \frac{1}{\sqrt{3}} & \frac{1}{\sqrt{2}}-a_6
\end{array}
\right)
$ 

\\&&
$
\left(\begin{array}{c}
m_0(a+b-d-\frac{\epsilon}{2}),\\
m_0(a+2d+\epsilon),\\
m_0(b+d-a+\frac{\epsilon}{2})
\end{array}
\right)
$
&
\\ \hline
\end{tabular}
\end{center}
\caption{
The mass matrix taking all three 
singlets, its mass eigenvalues
and its mixing matrix. The correction factors in the last row are:
$a_2=\frac{\sqrt{3} \epsilon}{4\sqrt{2} (a-d)}$, 
$a_3=-\frac{\sqrt{3} \epsilon}{4\sqrt{2} (a-d)}$ and 
$a_4=\frac{\epsilon}{2\sqrt{2} (a-d)}$, $a_5=\frac{1
\epsilon}{4\sqrt{2} (a-d)}$, $a_6=\frac{\epsilon}{4\sqrt{2} (a-d)}$.
}
\label{tab:threehiggs}
\end{table}

\begin{figure}[t]
\centering
\includegraphics[width = 19cm,height=12cm,angle = 0]{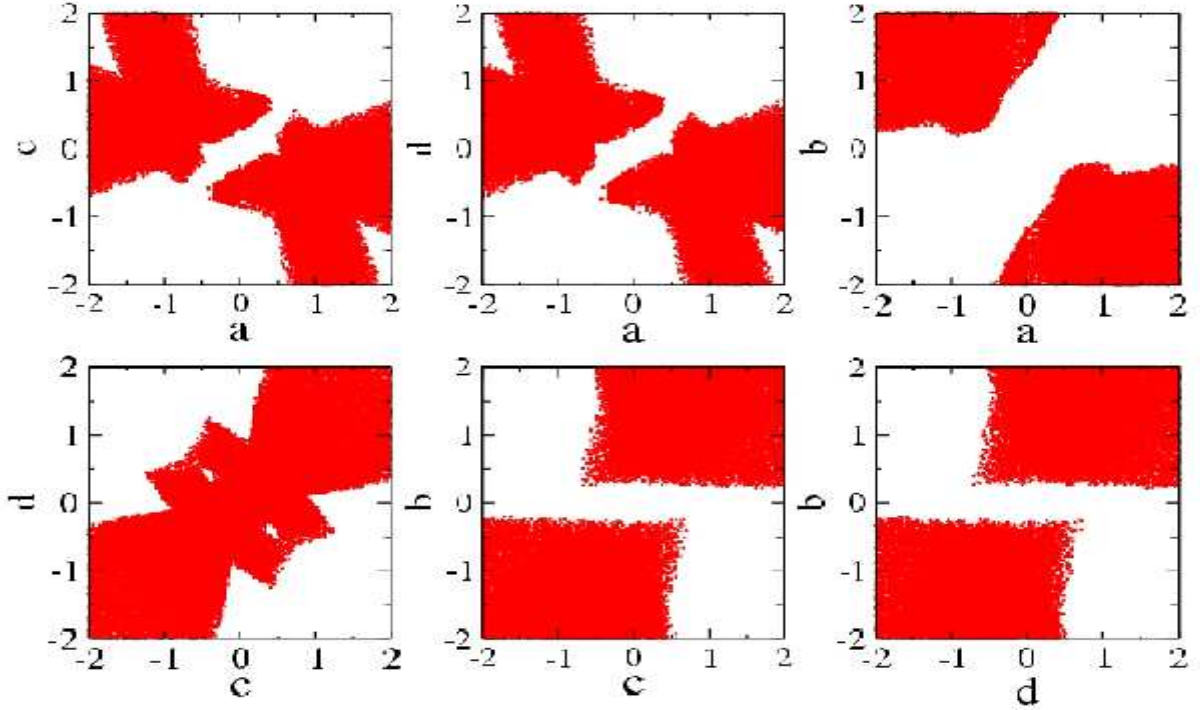}
\caption{Scatter plot showing the $3\sigma$ allowed regions in the
model parameter space for the case where 
$\xi$, $\xi^\prime$ and $\xi^{\prime \prime}$ 
all acquire VEVs.
The upper panels show allowed regions
projected onto the $a-c$, $a-d$, $a-b$ planes. 
The lower  panels show allowed regions
projected onto the $c-d$, $c-b$, $d-b$ planes. 
Normal hierarchy is assumed.}
\label{fig:threehiggs}
\end{figure}

\begin{figure}[t]
\centering
\includegraphics[width = 19cm,height=12cm,angle = 0]{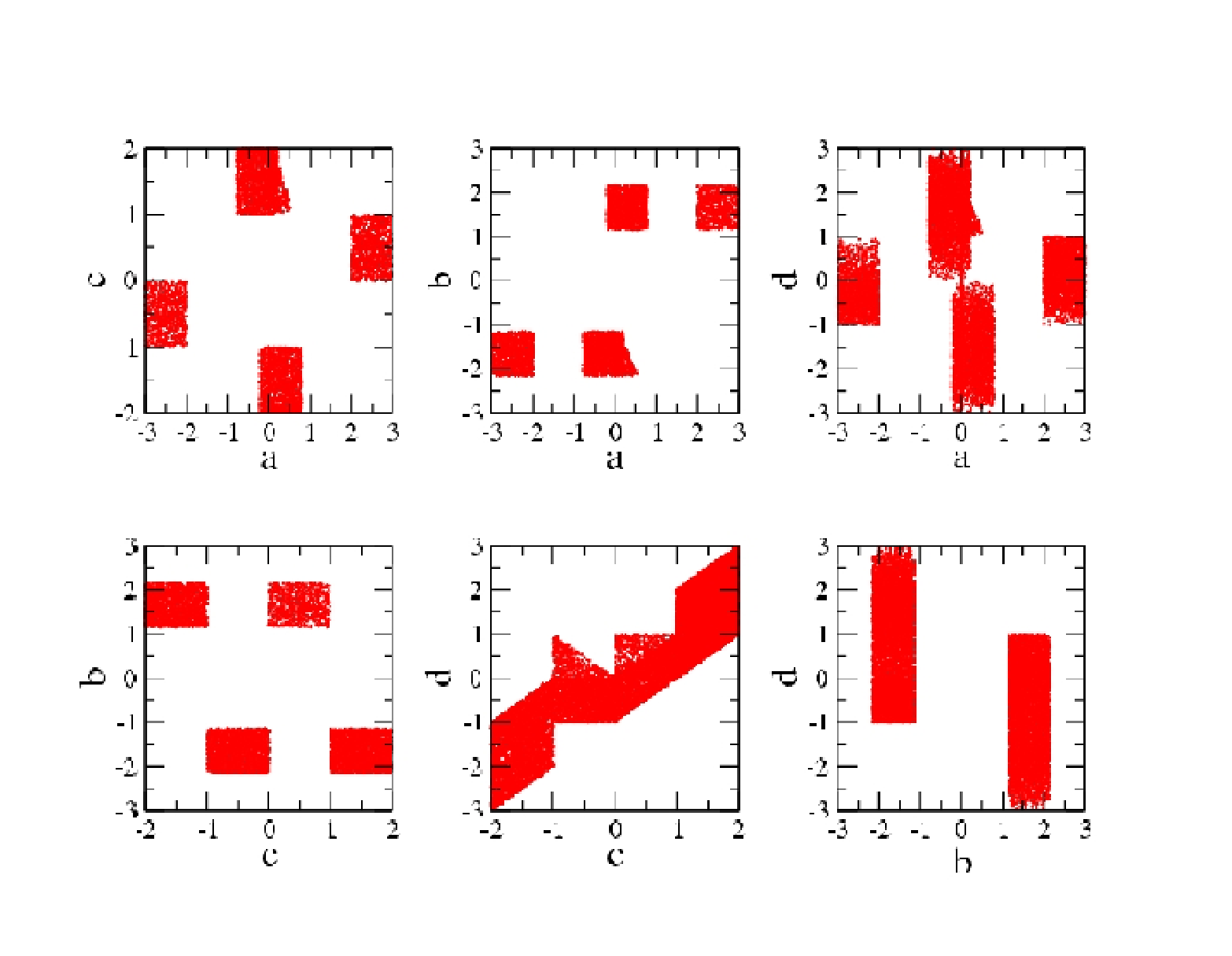}
\caption{\label{fig:threehiggsih}
Same as Fig. \ref{fig:threehiggs} but for inverted hierarchy.
}
\end{figure}

Finally, we let all three 
singlet Higgs VEVs contribute to \mnu. 
In this case one has to diagonalize the most general mass matrix 
given in Eq. (\ref{eq:phisalignedmnu}). This matrix has four independent 
parameters. If we assume that the singlet VEVs are such that 
$a=c=d$, \footnote{Note that one would need 
additional symmetries to explain $a=c=d$.}
then the eigenvalues and mixing matrix 
are given in the 
first row of Table \ref{tab:threehiggs}. 
This gives us a mass 
matrix whose structure is identical to that given in 
Eq. (\ref{eq:phisalignedmnu0}). Hence, it is not 
surprising that the corresponding mixing matrix that we 
obtain has exact TBM mixing and two of the mass eigenstates are 
degenerate. Therefore, to get the correct mass splitting 
it is essential that (i) we should have contribution from the 
singlet VEVs and (ii) the contribution from the 
the three singlets should be different. 
If we assume that $a = c \neq d$, then one can easily check that 
\mnu~ has $e-\tau$ exchange symmetry, and hence the 
resulting mass matrix is disallowed. This is because for 
$a=c$, as discussed before we get $e-\tau$ exchange symmetry 
and the $\xi^{\prime}$ term has an in-built
$e-\tau$ symmetry.  
Similarly for $a = d \neq c$, one gets $e-\mu$ 
symmetry in \mnu~ and is hence disfavored. 
Only when we impose the condition $c=d$, 
we have $\mu-\tau$ 
symmetry in \mnu, since the $\xi$ term and 
the sum of the $\xi^\prime$ and 
$\xi^{\prime\prime}$ terms are now separately 
$\mu-\tau$ symmetric. Therefore, the case 
$a \neq c = d$ gives us the TBM matrix and the mass eigenvalues 
are shown in Table \ref{tab:threehiggs}.
 
Since $a \neq c = d$ is the only allowed case for the three singlet 
Higgs case, we find the eigenvalues and the mixing 
matrix for the case where $c$ and $d$ are not equal, 
but differ by $\epsilon$. 
We take $c=d+\epsilon$ and for small values of $\epsilon$ 
give the results in the last row of 
Table \ref{tab:threehiggs}, keeping just the first order terms 
in $\epsilon$. 
The deviation from TBM is given as follows
\be
D_{12} \simeq 0,~~~~D_{23} \simeq \frac{\epsilon}{4(a-d)},~~~~
U_{e3} \simeq \frac{\epsilon}{2\sqrt{2}(a-d)}
~.
\label{eq:threehiggs}
\ee
The mass squared differences are
\be
\ms \simeq m_0^2\,(2a+b+d+\frac{\epsilon}{2})(3d-b+\frac{3\epsilon}{2})
,~~~
\ma \simeq  2\,m_0^2\,b(2d-2a+\epsilon)
~.
\label{eq:delmsq3higgs}
\ee
From the expression of the mass eigenvalues given in the Table, 
one can calculate the observables $m_t$, $m^2_\beta$ and 
$\langle m_{ee} \rangle$
\be
\langle m_{ee} \rangle  =  m_0\,(a+\frac{2b}{3}),~~
m_t = \sum_i |m_i| ,~~
m_\beta^2 \simeq m_0^2\,(a^2 + \frac{4ab}{3} + 
 \frac{2b^2}{3}+2d^2 - \frac{4bd}{3}+2d\epsilon -\frac{2b\epsilon}{3})
~.
\ee

\subsubsection{Normal Hierarchy}

Let us begin by 
restricting the neutrino masses 
to obey the condition $m_1 < m_2 \ll m_3$ and 
allow $a$, $b$, $c$ and $d$ to take {\it any} random 
value and find the regions in the $a$, $b$, 
$c$ and $d$ space that 
give $\ms$, $\ma$ and the mixing angles within 
their current $3\sigma$ allowed ranges. 
This is done by 
numerically diagonalizing \mnu. The results are shown 
as scatter plots 
in Fig. \ref{fig:threehiggs}. To help 
see the allowed zones better, 
we have projected the allowed points 
on the $a-c$, $a-d$ and $a-b$ plane shown in the upper panels, 
and $c-d$, $c-b$ and $d-b$ plane in the lower panels. 
There are several things one can note about the VEVs and hence 
the structure of the resultant \mnu
\begin{itemize}
\item $a=0$ is allowed, since this gives a 
\mnu~ which has contributions from 
$\xi'$ and $\xi''$, discussed in section 3.3, 
\item  $b=0$ is never allowed since $b$ is needed for TBM mixing 
as pointed out before,
\item $a=b$, $a = c$ and $a = d$ are never allowed,
\item $c=d$ is allowed and we can see from the lower left-hand panel 
how much deviation of $c$ from $d$ can be tolerated,
\item $c=0$ and $d=0$ can also be tolerated when
$a\neq 0$.
\end{itemize}
All these features are consistent with the results of Table 
\ref{tab:threehiggs}. 

\subsubsection{Inverted Hierarchy}

In this case too its possible to get inverted hierarchy. 
The corresponding 
values of the parameters of \mnu~ which allow this  
are shown as scatter plots in Fig. \ref{fig:threehiggsih}.
Here $m_0$ has been allowed to take any value, and 
we show the points projected on the $a-c$, $a-b$, $a-d$ 
plane in the upper panels and $c-b$, $c-d$, $b-d$ plane in the 
lower panels. Each of these points also satisfy the $3\sigma$ 
experimental bounds of Table \ref{tab:current}. Note that 
for $a=0$ we get the same regions in $b$, $c$ and $d$, as 
in Fig. \ref{fig:twohiggsih}. 

\section{Vacuum Alignment of the Triplet Higgs}

\begin{figure}[t]
\centering
\includegraphics[width = 10cm,angle = 0]{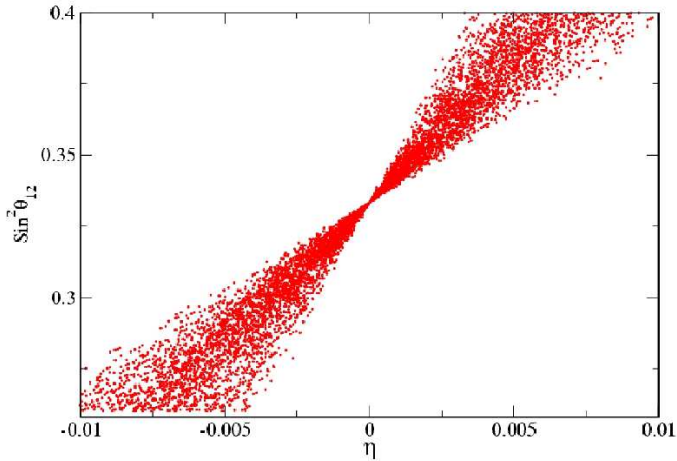}
\caption{\label{fig:vacalign}
Variation of $\sss$ with $\eta$ for the case where we allow 
for a misalignment of the triplet Higgs such that 
 $\langle \phi_S \rangle 
=(v_{S1},v_S,v_S)$ and $a \neq c=d$. 
}
\end{figure}

In case we do not confine ourselves to $\langle \phi_S \rangle 
=(v_S,v_S,v_S)$, we would have the general \mnu~ given in 
Eq. (\ref{eq:generalmnu}). Since we have argued 
in the previous section that the only viable scenario 
where one allows for all three Higgs singlet is 
when $a\neq c \simeq d$, 
we will assume that this 
condition for the singlet terms holds. We further 
realize that to reproduce a mixing matrix with 
$\theta_{13} \sim 0$ and $\theta_{23} \sim 45^\circ$, 
it might be desirable to keep $\mu-\tau$ symmetry in the 
mass matrix. Therefore, we show our results for 
the case $\langle \phi_S \rangle 
=(v_{S_1},v_S,v_S)$. The mass matrix is then given as
\be
{\cal M}_\nu = 
m_0
\pmatrix{
a+2b_1/3 &d-b/3& d-b/3 \cr
d-b/3  & d+2b/3 & a-b_1/3 \cr
d-b/3& a-b_1/3& d+2b/3 \cr
}
~.
\label{eq:alignmnu}
\ee
Of course for $b_1=b$ one would recover the case 
considered in the previous section and TBM mixing would 
result. The possibility of $b_1 \neq b$ gives rise to deviation 
from TBM mixing. 
In order to solve this matrix analytically we assume 
that $b_1 = b + \eta$ and keep only the first order 
terms in $\eta$. 
The mass eigenvalues obtained are
\be
m_1 = m_0\,(a+b-d + \frac{\eta}{3}),~~
m_2 = m_0\,(a+2d),~~
m_3 = m_0\,(-a+b+d + \frac{\eta}{3}),~~
~,
\ee
and the mixing matrix is
\be
\pmatrix{
\sqrt{2 \over 3}\bigg(1 - \frac{\eta}{3(3d-b)}\bigg)
& \sqrt{ 1 \over 3}\bigg(1+ \frac{2\eta}{3(3d-b)}\bigg)
& 0 \cr
-\sqrt{ 1 \over 6} \bigg(1 + \frac{2\eta}{3(3d-b)}\bigg)
& \sqrt{ 1 \over 3}\bigg(1 - \frac{\eta}{3(3d-b)}\bigg) & - \sqrt{ 1 \over 2} \cr
-\sqrt{ 1 \over 6}\bigg(1 + \frac{2\eta}{3(3d-b)}\bigg) 
& \sqrt{ 1 \over 3}\bigg(1 - \frac{\eta}{3(3d-b)}\bigg) & \sqrt{ 1 \over 2}
}.
\ee
Therefore, the only deviation from TBM comes in $\theta_{12}$ 
and we have
\be
D_{12} \simeq \frac{4\eta}{9(3d-b)}
~.
\ee
We show in Fig. \ref{fig:vacalign}
variation of $\sss$ with $\eta$. 
As expected, $\sss$ is seen 
to deviate further and further from its TBM value of 1/3
as we increase the difference between $v_S$ and $v_{S1}$. 
The other two mixing angles are predicted 
to be exactly at their TBM values due to the presence of  
$\mu-\tau$ symmetry in \mnu. They would also deviate from 
TBM once we allow for either $v_{S_2} \neq v_{S_3}$ or 
$c\neq d$, and in the most general case, both.

\section{Conclusions}

Current data seems to be pointing towards the existence of 
tribimaximal mixing. One needs to invoke some symmetry argument 
in order to get tribimaximal mixing. The discreet symmetry group 
$A_4$ has received a lot of attention in recent times as an 
attractive option for explaining the masses and mixing pattern of 
the neutrinos along with those of the charged leptons. 
We made a detailed phenomenological study of the viability 
of the different mass matrices that can be generated by 
spontaneous $A_4$ symmetry breaking. 

In particular, we considered the model proposed in \cite{ref5} and  
studied the phenomenological implications for it. The authors 
of \cite{ref5} consider only one $A_4$ 
singlet and one $A_4$ triplet Higgs contribution to 
the neutrino mass matrix. 
Since the singlet transforms as $1$ and since they 
take the vacuum alignment 
$\langle \phi_S \rangle = (v_S,v_S,v_S)$ for the $A_4$ triplet, 
the neutrino mass matrix by construction produces 
tribimaximal mixing. A lot of attention has been paid on 
justifying the vacuum alignment of the triplet Higgs which is 
absolutely essential for tribimaximal mixing. 
In this paper we pointed out that in addition to the 
vacuum alignment of the triplet, one also needs a certain 
fixed relation between the product of the VEV and the Yukawa of the 
singlet and the triplet. In particular, we found that in order to 
generate the correct ratio of $\ms$ to $\ma$, one demands that 
$b \simeq -1.88 a$, where $a=2 x_a\frac{u}{\Lambda}$ and 
$b=2 x_b \frac{v_{S}}{\Lambda}$. This appears to be extremely 
contrived and therefore undesirable. Even if one 
includes the $3\sigma$ uncertainties on $\ms$ and $\ma$, 
$|b|$ is fine tuned to $|a|$ within a factor of about $10^{-2}$.   
Even with this 
fine adjustment of the product of the VEVs and Yukawas, one 
would be able to generate only normal hierarchy for the neutrino
mass spectrum. 

We checked if it was possible to generate a viable mass matrix 
if the singlet Higgs belonged to either the $1'$ or $1''$
representation. We found them to be unsuitable 
due to the in-built wrong $S_2$ symmetry of the neutrino mass 
matrix for these cases. We studied the possibility of combining 
two singlet Higgs 
at a time. We showed that the 
case where $\xi'$ and $\xi''$ 
transforming as $1'$ and $1''$ 
are taken together is the only viable 
option, since this could lead to an approximate 
$\mu-\tau$ symmetry for the mass matrix. In the limit that 
$c=d$, we get exact $\mu-\tau$ 
symmetry and exact tribimaximal mixing. 
We allowed the 
breaking of this $\mu-\tau$ symmetry and showed how 
the mixing angles deviate from their tribimaximal values. We 
gave approximate analytical predictions for $\ms$, $\ma$, $m_t$ and 
$m_\beta^2$,  
when the difference between the 
model parameters $c$ and $d$ is small. We showed 
numerically through scatter plots how some of 
these quantities varied, as the difference between $c$ and $d$ 
was increased. We identified the regions of the 
model parameter space which produce values of the neutrino 
oscillation parameters within their current $3\sigma$ limits. 
This model is capable of producing even 
inverted hierarchy. We showed the regions of the $b-c-d$ 
parameter space which allow $m_3 < m_1 <m_2$ and at the same 
time correctly reproduce the neutrino oscillation parameters within 
their $3\sigma$ ranges. 

We next allowed for all three singlet Higgs contribution to 
the neutrino mass matrix. The case where $a \neq c=d$ emerged 
as the only possible case which produced exact tribimaximal mixing.   
Allowing $c\neq d$ allows for deviation from tribimaximal 
mixing. We studied this deviation as a function of the 
difference between $c$ and $d$. This model can give both normal 
and inverted hierarchy. The regions of the model parameter space which 
reproduce neutrino oscillations parameters within their 
current $3\sigma$ allowed ranges were identified both for the normal 
as well as for the inverted hierarchy. 

Finally, we allowed $\langle \phi_S \rangle$ to deviate from 
$(v_S,v_S,v_S)$. Changing the vacuum alignment immediately 
affects the tribimaximal character of the neutrino mixing 
matrix. We showed the results for 
$\langle \phi_S \rangle=(v_{S_1},v_S,v_S)$ and $a\neq c=d$ 
only, just for the 
sake of illustration. 
Since the resultant mass matrix by 
construction still has the residual $\mu-\tau$ symmetry, 
$\theta_{23}$ and $\theta_{13}$ remain at their 
tribimaximal values of $45^\circ$ and 0 respectively. 
We showed the deviation of $\sss$ from 1/3, analytically for  
small values of $\eta$ and numerically for all values of 
$\eta$, where $\eta$ quantifies the difference between $v_{S}$ and 
$v_{S_1}$. The most general case would of course be 
when $v_{S_1} \neq v_{S_2} \neq v_{S_3}$, where all 
three mixing angles will deviate from their tribimaximal 
values. Also, the condition $c=d$ is extremely ``unnatural'' and 
any deviation from that would also break tribimaximal mixing. 

\vglue 0.5cm
\noindent
{\Large{\bf Acknowledgments}}\vglue 0.3cm
\noindent
The authors wish to thank A. Raychaudhuri for very helpful 
discussions. M.M. thanks G. Altarelli for a communication. 
This work has been supported by the Neutrino Project
under the XI Plan of Harish-Chandra Research Institute.
B.B. would like to thank K. S. Babu for discussions.
The work of B.B. is supported by UGC, New Delhi, India, under
the Grant No F.PSU-075/05-06.


\end{document}